\documentstyle[eqsecnum,aps]{revtex}
\tightenlines

\def\bH{{\bf H}}
\def\bc{{\bf c}}
\def\ptl{\partial}
\def\br{{\bf r}}
\def\bx{{\bf x}}
\def\bk{{\bf k}}
\def\bn{{\bf n}}

\def\avg#1{\langle#1\rangle}
\def\bavg#1{\left\langle#1\right\rangle}

\def\al{\alpha}
\def\gam{\gamma}
\def\eps{\epsilon}
\def\dta{\delta}
\def\tta{\theta}
\def\Dta{\Delta}
\def\Om{\Omega}

\begin{document}
\draft

\title{Antiferromagnetism and Superconductivity in UPt$_3$}

\author{Anupam Garg}
\address{Department of Physics and Astronomy, Northwestern University,
Evanston, Illinois 60208}

\date{today}

\maketitle

\begin{abstract}

The short ranged antiferomagnetism recently seen in UPt$_3$ is proved incompatible
with two dimensional (2D) order parameter models that take the antiferromagnetism
as a symmetry breaking field. To adjust to the local moment direction, the order
parameter twists over very long length scales as per the Imry-Ma argument.  A
variational solution to the Ginzburg-Landau equations is used to study the nature
of the short range  order.  Although there are still two transitions, the lower
one is of first order --- in contradiction to experiments. It is shown that the
latent heat predicted by the 2D models at the lower transition is too large not to
have been seen.  A simple periodic model is numerically studied to show that
the second transition can not be a crossover either.
\end{abstract}

\widetext

\section{Introduction}
\label{Intro}

The purpose of this paper is to discuss the implications of recent experiments
\cite{Lus} on
antiferromagnetic (AFM) order in UPt$_3$ for the superconducting state. It is found 
that (i) a very weak AFM that sets in at 5 K is 
multi-domain in structure with a domain size of about $300$\AA, and (ii) despite a magnetic
moment per U atom of only $0.02\ \mu_B$, the domains can neither
be moved nor oriented by magnetic fields upto 3.2 T. We shall show that these findings pose
great difficulty for certain widely studied models \cite{msl,Sau}
in which the superconducting order parameter transforms as one
of the two dimensional (2D) represenations of the
relevant point group, $D'_{6h}$, and in which the AFM (or even less
plausibly, a structural modulation) provides a symmetry breaking field that splits the
superconducting transition into two just as an applied magnetic field splits
the A transition in superfluid $^3$He.  We believe that these data make it overwhelmingly
likely that superconducting UPt$_3$ is described by two order parameters belonging
to independent 
representations \cite{CG,ZU},
although the precise representations involved are still debatable.

Two main experimental facts with which any theory of UPt$_3$ must agree are: 
(A) There are two superconducting transitions at $T_{c+}$ and $T_{c-}$ in zero
field \cite{Fish},
$T_{c+}\simeq 550$~mK, and $\Delta T_c = T_{c+} - T_{c-} \simeq 50$~mK. Both
transitions appear to be of second order, with no observable latent heat.
(B) For basal-plane magnetic fields, ${\bH \perp \bc}$,
the phase diagram is isotropic~\cite{SRH}; i.e., it is the same
for all directions of field.\footnote{
We shall ignore an anisotropy of 4-5\% \cite{fn1} 
as it is too small to affect our arguments.}
We deliberately do not exploit other facts that also bear on the order parameter symmetry
in this paper. Chief among these is that for ${\bH \perp \bc}$,
there is a tetracritical point in the magnetic field-temperature ($H$-$T$) plane where
four second order lines meet \cite{Aden}.  The original 2D models \cite{msl} are
incompatible with a tetracritical point unless ${{\bf H}}\perp{\bf c}$ \cite{CG,Luk},
which also argues against them, as
the observed phase diagrams for ${{\bH}\| {\bc}}$ and ${\bf H}$ at $45^\circ$ to
$\bf c$ are similar to those for ${\bH \perp \bc}$ \cite{Aden,Lin},
although the tetracritical points in these cases may only be apparent
ones due to limited resolution of the phase boundaries \cite{ZU,Joy}.
Secondly, a specific $E_{2u}$ model {\it is} compatible with
a tetracritical point for all field orientations \cite{Sau}. It is therefore useful to have
independent arguments that constrain the order parameter without
relying on the presence or absence of tetracritical points.

With this motivation, we analyze in this paper the Ginzburg-Landau (GL) theory for 2D
models \cite{msl,Sau} when the AFM symmetry breaking field is allowed to vary randomly in
direction over a length scale
$a = 300$\AA.\footnote{
To avoid misunderstanding, we stress that
we are not studying the AFM transition itself, or the reason for AFM
disorder \cite{Tr}. Since the AFM order sets in well above
the superconducting transition, it is justified to treat it as an external field.
The spatial variation of this field in no way invalidates the usual argument for using
GL theory to interpret superconducting transitions.
Briefly, the argument (also known as the Ginzburg criterion) is that
the critical region is of width $(d/\xi_0)^4$ relative to $T_c$, where $d$ is an
inter-electron distance, and $\xi_0$ is the coherence length. For UPt$_3$,
$d\sim5$\AA, and
$\xi_0 \sim 200$\AA, so this region is unobservably small. In particular, the
question whether the transition is sharp or smeared out by the
random AFM is experimentally irrelevant.}
Since the problem involves an order parameter with a continuous symmetry in
the presence of a random field, the Imry-Ma argument~\cite{IM} implies that the
long-ranged average of the order parameter itself must vanish. Other forms of
order are not ruled out, however, and a balance of gradient and field orientation
energies shows~\cite{go4} that there are two transitions, the upper one of second
order, and the lower one of first order. If one is only interested in global
properties of the phase diagram and the universality class of the transitions, there
is nothing more to be said. We are more interested in whether
or not this scenario could describe the experimental data on UPt$_3$. To this end
we calculate the latent heat at the lower transition assuming that the coupling
to antiferromagnetism is weak. We find that this is too large to have been
missed by the experiments
\cite{Fish} if the splitting is taken to equal the observed value.
In other words, we can dismiss the possibility that the lower transition seen
experimentally is of first order with an unobservably small latent heat.
A second possibile explanation for the data
(which arises when the AFM coupling is not weak) is that
the two transitions predicted by the theory are too close
to be resolved, and that the lower specific heat jump is actually due to
a cross-over and not a true transition. We exclude this possibility too.
Using a simple periodic model for the AFM domain structure,
we find that the observed $H_{c2}$ slopes \cite{Aden,Lin}
and the {300~\AA} domain size imply that
a sufficiently sharp cross-over must be accompanied with a $\Delta T_c$ of order 80~K,
which is clearly absurd.

To orient the discussion further, let us review how
the 2D models account for facts A and B when the AFM is uniform.
The GL free energy $f_{\rm GL}$ for all these models can be written
in terms of a vector $\vec\eta = (\eta_x,\eta_y)$ with complex components.
In zero magnetic field $f_{\rm GL}$
is the sum of bulk, gradient,\footnote{
The gradient energy in the modified 2D model \cite{Sau} is slightly different, as
discussed below.}
and symmetry breaking field terms:
\begin{eqnarray} 
f_{{\rm bulk}} & = & -(T-T_0) \eta^*_i\eta_i + \beta_1(\eta_i^*\eta_i)^2
            +\beta_2|\eta_i\eta_i|^2, \label{fbulk} \\
f_{{\rm grad}} & = & \kappa_1\ptl_i\eta_j^*\ptl_i\eta_j +
    \kappa_2\ptl_i\eta_i^*\ptl_j\eta_j
               + \kappa_3\ptl_i\eta_j^*\ptl_j\eta_i \nonumber\\
              &&+ \kappa_4\ptl_z\eta_i^*\ptl_z\eta_i, \label{fgrad} \\
f_{{\rm SBF}} & = & 2\epsilon(|{{\bf n}}_i\eta_i|^2 - \eta_i^*\eta_i/2). \label{fsbf} 
\end{eqnarray}
Here, $i,j \in \{x,y\}$, and the summation convention is used, ${\bf n}$ is the
basal-plane unit vector along the magnetic moments, and $\epsilon$, $\beta_1$,
and $\beta_2$ are all positive. (If $\beta_2 < 0$, there is only one transition,
and the model can not describe UPt$_3$.) Suppose ${{\bf n}}=\hat y$ everywhere.
Then the first transition occurs at
$T_{c+}=T_0+\epsilon$ to a phase with $\vec\eta = e^{i\phi}(1,0)$. Ignoring the global
phase $\phi$, we refer to this as the real or (1,0) phase.
It breaks the rotational symmetry of the normal state. The second transition occurs at
$T_{c-} = T_0 - \beta_1\epsilon/\beta_2$ to a phase with
$\vec\eta = e^{i\phi}(1,iu)$ where $u$ is real and grows smoothly from 0 to
$\pm 1$ as $T$
is lowered further. This phase breaks rotational and time reversal symmetry,
and we shall refer to it as the axial or the $(1,i)$ phase. This explains fact A.
To explain fact B, it is necessary in the original 2D models \cite{msl}
to assume that the symmetry breaking field rotates in
response to the applied magnetic field. A symmetry breaking perturbation that was
locked to the lattice would split the $T_c$'s, but would not yield
an isotropic basal-plane phase diagram. In particular, the tetracritical point
would not exist for some $\bH$ orientations.~\footnote{
It follows that a structural modulation which should, a priori, be
insensitive to the orientation of $\bf H$, also can not be a
viable symmetry breaking field.}
This assumption was justified by hypothesizing \cite{go4}
that the AFM anisotropy energy was so weak that the N\'eel vector
reoriented itself to stay perpendicular to the magnetic field even for relatively
small $H$, and this idea has even been refined \cite{sau2} to explain the minute
anisotropy of $H_{c2}$ found by Keller et al. \cite{fn1}.

The neutron scattering data of Lussier et al. \cite{Lus} directly refute the
above hypothesis. They find that an AFM domain is of average size
$a\simeq 300\ $\AA\ \footnote{
The absence of anomalies in heat capacity~\cite{Fish2} or
NMR linewidth, Knight shift, or $1/T_1$~\cite{NMR},
raises doubts whether the AFM is static long range order; we set these doubts
aside here as they make the 2D models even more problematic.}
and is associated with one of three choices for the
spin density wavevector ${\bf q}$, with ${{\bf n}}\|\bf q$. The scattering
intensity under the peaks with different $\bf q$'s is essentially unchanged by fields
of up to 3.2 T, for either ${{\bH}}\|\bf q$ or $\bH\perp\bf q$, and in both field-cooled
and zero-field-cooled experiments. In addition, application of a field causes no
transfer of intensity to a scattering wave vector $\bf Q$ which is forbidden
(because $\bf Q\times\bf n=0$) when $H=0$.  This shows that the moments do not
cant away from the $\bf q$ direction while preserving the wavevectors.

Why, given these data, should one further investigate the 2D models? One
reason is the modified 2D model \cite{Sau}. In this model,
$\kappa_2=\kappa_3=0$, and the AFM is assumed to split the $\kappa_1$ term into
$\kappa_1^+|\ptl_i\eta_{\|}|^2 + \kappa_1^-|\ptl_i\eta_{\perp}|^2$,
where $\eta_{\|}$ and $\eta_{\perp}$ are components of $\vec\eta$ parallel and
perpendicular to $\bf n$, and $\kappa_1^+ - \kappa_1^- = {{\cal O}}(\epsilon)$.
(A similar splitting is assumed for $\kappa_4$.)
Note that the AFM continues to couple directly to the order parameter as
described by Eq.~(\ref{fsbf}). The linearized GL equations for
$\eta_{\|}$ and $\eta_{\perp}$ then decouple
even when $H\ne 0$, and an isotropic basal-plane phase diagram is recovered
even if the AFM is rigidly locked to the lattice. However, the random AFM excludes
this model too, as the scaling of gradient and field orientation energies
with $\eps$, and the overall balance between them,
is unchanged. A second reason is that although the general thrust of the
Imry-Ma argument \cite{IM} must continue to hold, the problem for UPt$_3$ is not
directly equivalent to an $n$-component vector spin model, and we are unaware of any
specific analysis for it. In particular we know of no careful comparison of the
energetics of the axial phase versus a real state with a wandering axis of orientation.

The rest of the paper is organized as follows. The GL free energy is analyzed in Sec. II.
We perform a variational analysis to understand the nature of the short-range order
at the upper transition, and also calculate the latent heat at the lower transition.
The simple periodic model is analysed in
Sec.~III. A brief discussion in Sec. IV concludes the paper.

\section{Analysis of the GL free energy}
\label{anal}

We wish to minimize the GL free energy, which is still given by 
Eqs.~(\ref{fbulk}--\ref{fsbf}), but ${{\bf n}}({\bf r})$ is now ${\bf r}$ dependent with
a correlation length $a=300\ $\AA. We first simplify the gradient energy to
\begin{equation}
f_{\rm grad} = \kappa |\nabla\eta_i|^2, \label{simgrad}
\end{equation}
as we are interested only in the competition between the gradient
and the symmetry breaking energies, and not in the precise slopes of the $H_{c2}$ lines
and whether they cross or not.\footnote
{This simplification is equally justified for the modified 2D model \cite{Sau}. The
gradient energy cost for variations of $\vec\eta$ along and perpendicular to $\bf n$ is now
slightly different, but it is, nevertheless, always positive. Any suitable average 
for the gradient energy scale $V_d$ therefore suffices for the scaling arguments
that follow.}
It is convenient to write ${{\bf n}}=(\cos\theta,\sin\theta)$, and introduce
the gradient energy scale $V_d = \kappa/a^2$, and circular polarization components,
$\eta_{L,R} = 2^{-1/2}(\eta_x \pm i\eta_y)$.

The first step is to find the critical temperature $T_c$, and the nature of the
critical order parameter. To do this we drop the quartic terms in $f_{\rm GL}$, and
minimize the rest. This yields an eigenvalue problem given by the linearized GL
equations:
\begin{eqnarray}
- \kappa\nabla^2\eta_L  + \epsilon e^{2i\theta}\eta_R  & = & -t\,\eta_L, \label{evL} \\
- \kappa\nabla^2\eta_R  + \epsilon e^{-2i\theta}\eta_L  & = & -t\,\eta_R, \label{evR}
\end{eqnarray}
where $t=T_c-T_0$ is the $T_c$ {\it enhancement}.
The critical solution is that with the {\it largest} value of $t$.

Two qualitatively different types of solutions to
Eqs.~(\ref{evL}) and (\ref{evR}) are as follows.
The first, or local solution, is motivated by the answer
for uniform $\theta$ \cite{msl}. Then $\vec\eta$
is also uniform, the gradient terms vanish, and we get
$\eta_L = - e^{2i\theta}\eta_R$, $t = \epsilon$.
If we now require $\vec\eta\perp{\bf n}$, i.e.,
$\eta_L=-e^{2i\theta}\eta_R$, even when $\theta$ is varying, then we pick up
a gradient energy of the order of $V_d$, and $t_{{\rm local}} = \epsilon - \al V_d$,
where $\alpha = 2 a^2 \langle (\nabla\theta)^2 \rangle$ is a number
of order one, and $\langle\cdot\rangle$ denotes a spatial average.
This solution does well only when $\epsilon \gg V_d$, i.e., when the symmetry
breaking energy is dominant.  The second, or real solution, has the form
$\vec\eta= e^{i\phi}\bf m$, $\bf m$ being a real unit vector varying on a length
$L\gg a$. The gradient energy is clearly of order $V_d(a/L)^2$. A random walk
argument for symmetry breaking energy leads to the estimate
$-\epsilon(a/L)^{3/2}$. Minimizing the sum of these two energies, we obtain
$L\sim a(V_d/\epsilon)^2$, and $t_{{\rm real}} \simeq \epsilon(\epsilon/V_d)^3$.
This
solution requires $V_d\gg \epsilon$ for consistency, and the $T_c$ enhancement is
strongly reduced compared to the uniform case.

A correct estimate of the symmetry breaking energy for the real solution
is more subtle, however. Although long range order in $\vec\eta$ is forbidden,
it is not obvious that the estimate of the energy itself is correct,
as one could try and lower it by short range adjustment to the local value of $\bn$.
Indeed, we will find that though the $T_c$ {\it splitting} is governed by the
$\eps^4/V_d^3$ energy gain, this gain is subdominant to that due to the
short-ranged order, and the {\it enhancement} in $T_c$ itself is larger, of order
$\eps^2/V_d$. 

The above results follow from a variational argument, which 
we present in three stages. In the first stage, let us consider 
the trial wave function $\eta_L = 1$, $\eta_R=\zeta e^{-2i\theta({\bf r})}$, 
with $\zeta$ as a variational parameter. We call this the axial solution. (See
reasons below.) The optimal value of $\zeta$ is to be found by minimizing
\begin{equation}
W = {\langle\eta|H_{GL}| \eta\rangle \over \langle\eta|\eta\rangle},
\label{Wdef}
\end{equation}
where $H_{GL}$ is the linear operator, or GL ``Hamiltonian" in
Eq.~(\ref{evL}) and (\ref{evR}), $|\eta\rangle$ is the trial solution, and we use an
obvious quantum mechanical notation for the averages.
A simple calculation gives
\begin{equation}
W = 2{\eps\,{\rm Re}(\zeta) + \al V_d |\zeta|^2 \over 1 + |\zeta|^2}.
\label{Wsimp}
\end{equation}
[The quantity $\alpha = 2 a^2 \langle (\nabla\theta)^2 \rangle$ as before.]
Minimization gives a $T_c$ enhancement
$t_{\rm axial}  =  -\alpha V_d + (\epsilon^2 +\alpha^2 V_d^2)^{1/2}$,
which is $ \approx\epsilon^2/2\alpha V_d$ for $\epsilon\ll V_d$, and
$\approx \epsilon - \alpha V_d$ when $\epsilon \gg V_d$.
We have thus done better than both the local and the real solutions.

In the second stage, we generalize the above solution by taking
\begin{equation}
\eta_L = 1,\ \ \eta_R(\br) = \int d^3\br' Q(\br - \br')e^{-2i\theta(\br')}
     \equiv h(\br), \label{etvar}
\end {equation}
where $Q(\br)$ is a variational kernel. We take
$e^{2i\theta(\br)}$ to be Gaussian-distributed, with
$\langle e^{2i\theta(\br)}\rangle = 0$, and a correlation function
$g(\bx) = \langle e^{2i\theta(\br)}e^{-2i\theta(\br+\bx)}\rangle$ decaying on a length
scale $a$. Denoting the volume of the system by $\Om$, the average of the
symmetry breaking energy is found to be
\begin{equation}
2\ {\rm Re}{\eps \over \Om}\int d^3r\ \eta_L^* \eta_R e^{2i\theta(\br)}
    = 2\eps F_1[Q],
\label{esbf}
\end{equation}
where $F_1[Q]$ is the functional
\begin{equation}
F_1[Q] = \int {d^3k \over (2\pi)^3} Q(\bk) g(\bk),
\label{F1}
\end{equation}
where $Q(\bk)$ and $g(\bk)$ are Fourier transforms of $Q(\br)$ and $g(\br)$. 
Similarly, the average of the gradient energy equals
\begin{equation}
{\kappa\over\Om}\int d^3r |\nabla\eta_R|^2 = \kappa F_2[Q],
\label{egrad}
\end{equation}
where
\begin{equation}
F_2[Q] = \int {d^3k \over (2\pi)^3} k^2 Q^2(\bk) g(\bk).
\label{F2}
\end{equation}
Finally, the spatial average of $|\eta_R|^2$, needed for normalization, is given by
\begin{equation}
{1\over\Om}\int d^3r |\eta_R|^2 = F_3[Q] \equiv
          \int {d^3k \over (2\pi)^3} Q^2(\bk) g(\bk).
\label{F3}
\end{equation}
The expectation of $H_{GL}$ is therefore given by
\begin{equation}
W = (2\epsilon F_1[Q] + \kappa F_2[Q])/(1+F_3[Q]), \label{hglav}
\end{equation}
which must be minimized with respect to $Q(\bk)$. Setting the variation of $W$
to zero, we obtain
\begin{equation}
(1+F_3)(2\eps\,\dta F_1 + \kappa\,\dta F_2) - (2\eps F_1 + \kappa F_2)\dta F_3 = 0.
\label{varcon}
\end{equation}
Using $\dta F_i/\dta Q(\bk) = g(\bk)$, $2 k^2 Q(\bk)g(\bk)$, and
$2 Q(\bk) g(\bk)$ for $i=1$, $2$, and~$3$, respectively,
and cancelling $g(\bk)$ in this equation, we get
\begin{equation}
(1+F_3)[\epsilon + \kappa k^2Q(\bk)] - (2\epsilon F_1 + \kappa F_2)Q(\bk) = 0.
\label{mincond}
\end{equation}
This yields the Yukawa-like form
\begin{equation}
Q(\bk) = - {\eps \over\kappa (k^2 + k_0^2)},
\label{Yukawa}
\end{equation}
where $k_0^2$ is self-consistently given by
\begin{equation}
k_0^2 = -{2\eps F_1 + \kappa F_2 \over \kappa (1+F_3)}
      = -{W_{\rm min} \over \kappa}.
\label{k0}
\end{equation}
In the last form, which follows by comparison with Eq.~(\ref{hglav}), $W_{\rm min}$
denotes the minimum value of $W$.

It remains to find the $F_i[Q]$ with the form (\ref{Yukawa}) and substitute in
Eq.~(\ref{k0}) to obtain $k_0$ and $W_{\rm min}$. Since $g(x)$ decays on a length
scale $a$, we expect $g(k)$ to decay on a scale $a^{-1}$. Since $g(\bx = 0) =1$, 
this decay must be faster than $1/k^3$ as $k\to\infty$; this fact will guarantee
convergence of the integrals below. For $F_1$ we obtain
\begin{equation}
F_1 = -{\eps\over\kappa}{1\over 2\pi^2} \int_0^{\infty}
      {1\over k^2 + k_0^2} g(k) k^2 dk.
\label{F1f}
\end{equation} Now, by the second equality in Eq.~(\ref{k0}), we
expect $k_0$ to vanish as $\eps \to 0$. It is therefore valid to put
$k_0$ to zero in the integrand above. This yields
\begin{equation}
F_1 \approx -c_1 (\eps/ V_d),
\label{F1ans}
\end{equation}
where
\begin{equation}
c_1 \equiv {1\over 2\pi^2 a^2} \int_0^{\infty} g(k) dk
\label{c1def}
\end{equation}
is a dimensionless number of order unity. In the same way, we get
\begin{equation}
F_2 = {\eps^2\over\kappa^2} {1\over 2\pi^2} \int_0^{\infty}
      {k^2\over (k^2 + k_0^2)^2} g(k) k^2 dk
      \approx c_1 (\eps a /\kappa)^2,
\label{F2ans}
\end{equation}
where the last result follows from again setting $k_0 = 0$ in the integrand.
Lastly, $F_3$ is given by
\begin{equation}
F_3 = {\eps^2\over\kappa^2} {1\over 2\pi^2} \int_0^{\infty}
      {1 \over (k^2 + k_0^2)^2} g(k) k^2 dk.
\label{F3f}
\end{equation}
It is impermissible to set $k_0 = 0$ now as the resulting integral is divergent.
Instead, we note that the integrand is sharply peaked near $k=0$, and use this fact
to replace $g(k)$ by $g_0 \equiv g(\bk = 0)$. This yields
\begin{equation}
F_3 \approx c_2 {1\over k_0 a} \left( {\eps \over V_d} \right)^2  
\label{F3ans}
\end{equation}
where $c_2 = g_0/8\pi a^3$ is another dimensionless number of order unity.

We will show shortly that $F_3 \ll 1$. (This is also to be expected from the
fact that $F_3 = \avg{|\eta_R|^2}$.) Assuming this to be the case, using
Eqs.~(\ref{F1ans}) and (\ref{F2ans}), and neglecting
$F_3$ in comparison with unity in Eq.~(\ref{k0}), we obtain
\begin{equation}
k_0 a = \sqrt{c_1} (\eps/V_d), \label{k0ans} 
\end{equation}
\begin{equation}
t_{\rm axial} = -W_{\rm min} =  c_1\eps^2/V_d.  \label{Wans}
\end{equation}
It then follows that $F_3 \sim \eps/V_d \ll 1$ self-consistently. The assumption
$k_0 a \ll 1$ is also seen to be true. Although the $T_c$ enhancement is
still of order $\eps^2/V_d$, we can see that it exceeds
that obtained with the earlier simple trial solution by noting
that the quantity $2 \al c_1$ can be written as the product $\avg{k^2} \avg{k^{-2}}$,
where $\avg{\cdot}$ is a mean defined with respect to the
distribution $g(\bk) d^3k/(2\pi)^3$. The desired result then follows from
noting that $2\al c_1 > 1$, i.e., $c_1 > 1/2\al$, either as a consequence of 
H\"older's inequality, or by recalling that the arithmetic mean of
positive numbers with a non-zero spread always exceeds the
harmonic mean. Further, the solution is
non-perturbative since $|\eta_R|\sim (\epsilon/V_d)^{1/2}$, which is not an integer
power of $\eps$.

The improvement over the real solution is illusory, however.
First, note that the axial solution is degenerate, since if $(\eta_L,\eta_R) = (u,v)$
is a solution to Eqs.~(\ref{evL}) and (\ref{evR}), so is $(v^*,u^*)$.
(This is merely complex conjugation.) Since $\eta_R \ne \eta^*_L$ for Eq.~(\ref{etvar}),
these are distinct solutions. On the other hand, the GL equations are seen to be
completely real when written in terms of $\eta_x$ and $\eta_y$, suggesting that the
ground state is also real and unique.\footnote{
Another way of saying this is that in a large but finite system, $H_{GL}$
has no symmetries (including time-reversal) with a random AFM. Thus this
argument does not apply to the periodic model studied in in Sec. III.}
Thus a linear combination of the
solution just found with its complex conjugate might be even lower in energy. The
second point is that the length scale $k_0^{-1}\sim \eps^{-1}$ is much less than the
Imry-Ma length scale $\sim \eps^{-2}$. Thus the axial order found above is only
short-ranged, and we can multiply the variational answer by a phase factor 
$\gam(\br)$ varying
on a length scale greater than $k_0^{-1}$ and try to further reduce the energy
in this way. This forms the third stage of our argument.
We denote the solution (\ref{etvar}) by ${\vec\eta}^{\,(1)}$, its
complex conjugate by ${\vec\eta}^{\,(2)}$, and examine the combination
\begin{equation}
{\vec u}(\br) = {i\over 2} \left( e^{i\gam(\br)}{\vec\eta}^{\,(1)}
                 - e^{-i\gam(\br)}{\vec\eta}^{\,(2)} \right).
\label{uvar}
\end{equation}
This has cartesian components
\begin{eqnarray}
u_x & = & - \sin\gam\,(1 +{\rm Re}\,h) - \cos\gam\,{\rm Im}\,h, \label{ux}\\
u_y & = &  \cos\gam\,(1 - {\rm Re}\,h) + \sin\gam\,{\rm Im}\,h.  \label{uy}
\end{eqnarray}
Since $|h(\br)| \ll 1$, this solution is close to $(-\sin\gam, \cos\gam)$,
i.e., to a rotation of the $(1,0)$
solution.  If $\gam$ were a constant, the energy would be identical to that of
the solution (\ref{etvar}), i.e., given by Eq.~(\ref{Wans}). To calculate
the difference, let us define
\begin{equation}
T_1 = T_0 + t_{\rm axial}, \label{T1}
\end{equation}
and $A^2 = 1 + \avg{|h(\br)|^2} = 1 + F_3$. If we average the quadratic part of $f_{GL}$
over a length scale of order $k_0^{-1}$, we obtain a coarse grained free energy
\begin{equation}
f_{cg} \approx A^2 \left( (T_1 - T) + \kappa (\nabla\gam)^2
        - \eps \bavg{\cos 2[\tta(\br) - \gam(\br)]}_{cg} \right).
\label{fcg}
\end{equation}
where $\avg{\cdot}_{cg}$ denotes a coarse grained average. Note that the first term
is proportional to $T_1 - T$ and not $T_0 - T$ as in Eq.~(\ref{fbulk}).
We can now apply the
Imry-Ma argument to Eq.~(\ref{fcg}). Suppose the energy is minimized when $\gam$
varies on a length scale $L$. In a supercell of linear size $L$, there are approximately
$(L/a)^3$ cells in each of which $\tta(\br)$ can be regarded as constant and independent
of the other cells. Thus the coarse grained average of either $\cos 2\tta$
or $\sin 2\tta$ in a supercell is a number of order $(L/a)^{3/2}$ of either sign.
By choosing $\cos 2\gam \propto +\avg{\cos 2\tta}_{cg}$, 
and $\sin 2\gam \propto +\avg{\sin 2\tta}_{cg}$, we obtain a symmetry breaking energy
which is systematically negative in every supercell. Since the gradient energy is
of order $L^{-2}$, we recover the Imry-Ma length scale $L\sim \eps^{-2}$, and an
energy lowering of ${\cal O}(\eps^4)$. This energy 
is {\it subdominant} to the gain (\ref{Wans})
from the short ranged order of Eq.~(\ref{etvar}), but since
it represents an additional gain, the true solution is indeed disordered and real.
It is in this sense that the real solution must be understood.

Therefore, the transition temperature for the real solution is given by
\begin{equation}
T_{c+} = T_1 + {\cal O}(\eps^4/V_d^3). \label{Tc+}
\end{equation}
(Actually, $T_1 - T_0$ could also contain
a term of order $\eps^4$, but that is immaterial.) Let us now consider what happens
at lower temperatures. Now the quartic terms come into play, and the positive
$\beta_2$ term in $f_{GL}$ favors an axial solution. For $\eps \ll V_d$, we can compare
the free energies of the real and axial solutions directly. Since
$\avg{|\eta_i\eta_i|^2}$ vanishes for a purely axial solution and equals
$\avg{(\eta_i\eta_i^*)^2}$ for a purely real one, these free energies will differ
strongly in the contribution of the $\beta_2$ term. Approximating multiplicative factors
such as $\avg{|\eta|^4} / \avg{|\eta|^2}^2$ by unity, and denoting real and axial solution
quantitites by subscripts $r$ and $a$, the standard analysis of
a Landau free energy with a second order transition yields
\begin{eqnarray}
f_r & = & - C_r (T - T_{c+})^2 /2 T_{c+}, \label{fr}\\
f_a & = & - C_a (T - T_1)^2/2 T_1, \label{fa}
\end{eqnarray}
where
\begin{eqnarray}
C_r/T_{c+} & = & 1/2(\beta_1 + \beta_2),\label{Cr}\\
C_a/T_1    & = & 1/2\beta_1.\label{Ca}
\end{eqnarray}
The $C$'s are specific heats that would be obtained for the respective solutions.
Since $C_a > C_r$, the free energies will intersect at a temperature $T_{c-} < T_1$
(see Fig. 1), giving a first order transition at $T_{c-}$. Ignoring the small
differences between $T_0$, $T_{c+}$, and $T_1$ compared to $T_0$ itself, we obtain
\begin{equation}
\Dta T_c = T_{c+} - T_{c-} = {\mu \over \mu -1}(T_{c+} - T_1), \label{dtc}
\end{equation}
where
\begin{equation}
\mu = (C_a/C_r)^{1/2}. \label{mu}
\end{equation}
Differentiation of the free energies yields the entropies $S_r$ and $S_a$,
and the latent heat $\ell = T(S_r - S_a)$:
\begin{equation}
\ell = (\mu - 1) C_r \Dta T_c. \label{latheat}
\end{equation}

Let us ask if the scenario that has emerged could describe the actual data.
To this end, we take $\Dta T_c$ to be the observed $T_c$ splitting
and $C_r$ and $C_a$ to be the specific heat jumps at the upper and lower transitions 
(relative to the normal state). Using data for sample 1 from Ref. \onlinecite{Fish},
$\Dta T_c =60$ mK, $C_r = 98$, and $C_a =113$ mJ/K~mole UPt$_3$, which implies
$\mu = 1.073$. Equation (\ref{latheat}) then gives $\ell = 0.43$ mJ/mole UPt$_3$.
If we conservatively take the experimental
temperature resolution to be 10 $\mu$K, then since the base specific heat is about
0.1 J/K~mole, we conclude that a latent heat of 1 $\mu$J/mole would have been detected.
The latent heat predicted by the 2D models is much larger.
Further, the measured \cite{Aden} $H_{c2}$ slopes of $\sim 5\ $T/K imply values
$\kappa=6.6\times 10^3\ $K\AA$^2$ and $V_d = 73\ $mK. To obtain a $\Dta T_c$
of 60 mK
we would therefore need $\eps \simeq 70$ mK, which is not much smaller than $V_d$.

\section{Simple periodic model}
\label{spm}

Tha analysis so far has been done assuming $\eps \ll V_d$. 
It is also interesting to ask what happens when $\eps \gg V_d$.
In this case we expect to obtain the local solution to a first approximation.
Suppose the solution is of the strictly local type near $T_{c+}$. Consider two
nearby domains with $\bn\|\bx$, and ${\vec\eta}\|{\bf y}$, and an intervening
domain with $\bn$ at 120$^\circ$ to ${\bf y}$ and ${\vec\eta}\perp\bn$. The system will
lower its gradient energy by tunneling of the $\eta_y$ component through the intervening
domain. Since the $\vec\eta$ component parallel to $\bn$ is exponentially small in the
ratio $\epsilon/V_d$, we
expect the magnitude of the gradient energy gain will also be similarly small.
However, the
gradient energy favors relatively real $\eta_x$ and $\eta_y$, while the $\beta_2$
term favors a relative phase of $\pi/2$ between them. General symmetry and
continuity arguments would suggest that we still have two phase transitions with the
lower one of first-order, but now the $T_c$-splitting, and the degree of
axiality at the lower transition will be exponentially small. It is then conceivable
that an appreciably axiality only develops as a crossover at a lower temperature.
This raises the possibility that the two true phase transitions are so close together
that they appear to be one experimentally, and the crossover appears to be the
phase transition at a lower temperature.

To investigate the above possibility, we have numerically studied a toy model with a
periodic antiferromagnetic domain structure. The assumption of periodicity
eliminates the possibility of a first order transition altogether. This is because
the model has an extra symmetry, and so is not in the same strict universality class
as that studied in Sec. II. While the periodic model can not capture the two very
close-by transitions (it only yields one), it {\it can} capture the low temperature
crossover. Since we are only interested in the latter behaviour in this section, the
periodic model is adequate. At the same time the periodicity simplifies the numerical
problem enormously. We further take ${\bf n}({\bf r})$ to vary only in one
direction ($x$), and alternate between {\it two} orientations $\hat{\bf x}$ and 
$\hat{\bf y}$ with a periodicity of $2a$. More concretely, we have
\begin{equation}
F_{\rm SBF} = \int dx\, 2\epsilon(x)(|\eta_x|^2 - |\eta_y|^2), \label{fspm}
\end{equation}
where $\epsilon(x) = \pm\epsilon$ for $ 0 < \pm x <  a$, and
$\epsilon(x) = \epsilon(x+2a)$. 
The restrictions of one-dimensonality and two orientations for $\bf n$ 
simplify the problem without changing its essential aspects. A closer analog
of the modified 2D model \cite{Sau}
would be obtained by allowing $\kappa$ to vary periodically in addition to $\eps$,
but since the fit to $H_{c2}$ data requires the two values of $\kappa$ to be
rather close, the gradient energies are not appreciably different from domain to
domain, and nothing is gained from this refinement.

The eigenvalue problem for $T_c$ is identical to the Kronig-Penney model, and
is analytically solvable. The solution is doubly degenerate corresponding to
either $\eta_x$ or $\eta_y$ being zero everywhere. (Note that with $V_d = \kappa/a^2$
as before, $T_c-T_0 \approx \epsilon^2/12 V_d$ for small $\epsilon$,
so that even with $\epsilon\simeq 3V_d$, say, one is not in the strong
symmetry breaking regime.) We solve the nonlinear GL equation below
$T_c$ by discretizing it with a mesh size $h\ge 2^{-10}a$, and using a Newton-Raphson
method. The free energy is computed for a closely spaced set of $T$ values
and numerically differentiated to obtain $C_v/T$. The calculations
are done with $\beta_1 = 0.5V_d$, and $\beta_2 = 0.35V_d$. This ratio of $\beta_2$
to $\beta_1$ is chosen to yield comparable $C_v/T$ jumps at $T_{c\pm}$
in the uniform case.
Our results for $C_v/T$ are shown in Fig. 2 for various $\epsilon/V_d$. There
is clearly no second transition.\footnote{
This conclusion is supported by analytic work for $\epsilon\ll V_d$,
in which $T_c$
is found by keeping only one wavevector $q=\pi/a$. A linear analysis
about this solution below $T_c$, and in the same approximation of one $q$, yields
no further instabilities, {\it except} to the complex conjugate solution at $T_c$
itself.}
For $\epsilon/V_d = 500$ there is
a crossover that could at first sight mimic the data of Ref. \onlinecite{Fish},
but this is not so.
As noted earlier, the measured \cite{Aden} $H_{c2}$ slopes imply $V_d = 73\ $mK.
Since the
temperature axis in Fig. 2 has been scaled by $\epsilon$, and $\epsilon = 36\ $K,
the implied $T_c$ splitting would be $2.3\epsilon\simeq 80\ $K. This is so far off the
measured $50\ $mK splitting
that the cross-over scenario within a 2D model can not be entertained for UPt$_3$.

The surprising result from this exercise is that one might have expected a crossover
and an apparent $T_c$ splitting of order $\epsilon$ to appear for
$\epsilon = 20 V_d$, say. This is not so, and the crossover is only manifest at a much
larger value of $\epsilon/V_d$.

\section{Discussion}

We have considered the effect of the short ranged antiferromagnetic domain
structure on the superconducting transition in UPt$_3$ within the 2D models, wherein
the antiferromagnetism is the agency that splits the transition.
It is known~\cite{go4} that when the AFM is weak and random, one does obtain
two transitions but the lower one is of first order.
We have shown that this prediction is in {\it quantitative} conflict with
observations by calculating the latent heat at this transition, and showing
that if the $T_c$ splitting and the specific heats are to agree with the observed data,
then this latent heat is well above experimental limits on detectability. In the
course of this analysis we have studied a variational axial state to
better understand the short-ranged order. We have also
studied the possibility that the lower transition may be a cross-over and not a true
phase transition. In this case, the observed $T_c$ splitting can not be reconciled
with the gradient energy scale. In either case, the AFM domain structure rules out
all 2D models \cite{msl,Sau}.

Finally, we briefly disuss experiments which show that the two transitions in UPt$_3$
merge into one at a pressure of $p_c \sim 4$~kbar \cite{HvL}.  The loose
interpretation that this happens because the antiferromagnetism disappears at about the
same pressure is thermodynamically unsustainable \cite{CG}. All mean field models,
whether based on one \cite{Joy} or two \cite{CG,ZU} primary order parameers,
require either the reemergence of a splitting for $p>p_c$, or
a first order line at $p\simeq p_c$. Neither possibility has been ruled out so
far in our view, as the broad heat capacity anomalies seen in \cite{HvL} could
conceal two transitions, and a first order transition has not
been looked for. Careful experimental study of these possibilities is clearly
highly desirable. At present, we believe that the two-order-parameter models,
especially the AB model \cite{CG}, are the only viable ones left for describing UPt$_3$.

\acknowledgments

Some of this work was done while visiting the Universit\"at Augsburg, Germany, and the
Institute for Scientific Interchange, Torino, Italy, and I thank my hosts at both
institutions for their hospitality. I am indebted to Rajiv R. P. Singh and David
Thouless for instructive discussions and correspondence about the Imry-Ma argument.

\begin{figure}
\caption{Sketch of the free energies (relative to that of the normal state)
of the real and axial solutions, showing a
change in slope, i.e., a first order transition at $T_{c-}$. The transition at
$T_{c+}$ is of second order.}
\end{figure}

\begin{figure}
\caption{Computed specific heat for the simple periodic model. $\epsilon$ and $V_d$ are
the symmetry breaking field and the gradient energies.}
\end{figure}


\begin{references}
\bibitem[*]{byline} Electronic address: agarg@nwu.edu

\bibitem{Lus}B. Lussier, L. Taillefer, W.J. L. Buyers, T. E. Mason, and
T. Peterson, Phys.\ Rev.\ B {\bf 54}, 6873R (1996).

\bibitem{msl}M. Ozaki, Prog.\ Theor.\ Phys.\ $\bf 76$, 1008 (1986); M. Sigrist, R. Joynt,
and T. M. Rice, Phys.\ Rev.\ B $\bf 36$, 5186 (1987);
G. Volovik, Zh.\ Eksp.\ Theor.\ Fiz.\ $\bf 48$, 39 (1988) [JETP Lett.\
{\bf 48}, 41 (1988); R. Joynt, Supercond.\ Sci.\ Technol.\ {bf 1}, 210 (1988);
D. W. Hess, T. Tokoyasu, and J. A. Sauls, J.\ Phys.\ Condens.\ Matter
{\bf 1}, 8135 (1989).

\bibitem{Sau}J. A. Sauls, Adv.\ Phys.\ {\bf 43}, 113 (1994).

\bibitem{CG}D. C. Chen and Anupam Garg, Phys.\ Rev.\ Lett.\ {\bf 70}, 689 (1993);
Phys.\ Rev.\ B 53, 374 (1996); Anupam Garg and D. C. Chen, Phys. Rev. B 49, 479 (1994).

\bibitem{ZU}M. E. Zhitomirsky and K. Ueda, Phys.\ Rev.\ B {\bf 53}, 6591 (1996).

\bibitem{Fish}R. A. Fisher et al., Phys.\ Rev.\ Lett.\ {\bf 62}, 1411 (1989).

\bibitem{SRH}B. S. Shivaram, T. F. Rosenbaum, and D. G. Hinks, Phys.\ Rev.\
Lett.\ 57, 1259 (1986);
L. Taillefer et al., J.\ Magn.\ Magn.\ Mater.\ {\bf 90-91}, 623 (1990).

\bibitem{fn1} N.  Keller et al., Phys.\  Rev.\ Lett.\ {\bf 73}, 2364 (1994);
{\bf 74}, 2148(E) (1995).

\bibitem{Aden}S. Adenwalla et al., Phys.\ Rev.\ Lett.\ {\bf 65}, 2298 (1990);
G. Bruls et al., {\it ibid} {\bf 65}, 2294 (1990).

\bibitem{Luk}I. Luk'yanchuk, J.\ Phys.\ I (France) {\bf 1}, 115 (1991);
I. Luk'yanchuk, M. Sigrist, and M. E. Zhitomirsky,
Phys.\ Rev.\ Lett.\ {\bf 71}, 1957 (1993).

\bibitem{Lin}S. W. Lin et al., Phys. Rev\ B {\textbf 49}, 10001 (1994).

\bibitem{Joy}R. Joynt, Phys.\ Rev.\ Lett.\ {\bf 71}, 3015 (1993);
K. A. Parks and R. Joynt, ibid {\bf 74}, 7734 (1995).

\bibitem{Tr} W. Trinkl, S. Cors\`epius, E. Guha, G. R. Stewart,
Europhys.\ Lett.\ {\bf 35}, 207 (1996).

\bibitem{IM}Y. Imry and S.-K. Ma, Phys.\ Rev.\ Lett.\ {\bf 35}, 1399 (1975).

\bibitem{go4}R. Joynt, V. P. Mineev, G. E. Volovik, M. Zhitomirsky, Phys.\ Rev.\
B {\bf 42}, 2014 (1990).

\bibitem{sau2} J. A. Sauls, Phys.\ Rev.\ B {\bf 53}, 8543 (1996).

\bibitem{Fish2}R. A. Fisher et al., Sol.\ St.\ Commun.\ {\bf 80}, 263 (1991).

\bibitem{NMR} Y. Kohori et al.,
J.\ Magn.\ Magn.\ Mater.\ {\bf 90-91}, 510 (1992); Physica B
{\bf 165-166} 381 (1990); Moohee Lee et al. Phys.\ Rev.\ B {\bf 48}, 7392 (1993).

\bibitem{HvL}H. v. L\"ohneysen, T. Trappmann, and L. Taillefer,
J.\ Magn.\ Magn.\ Mater.\ {\bf 108}, 49 (1992).

\end{references}
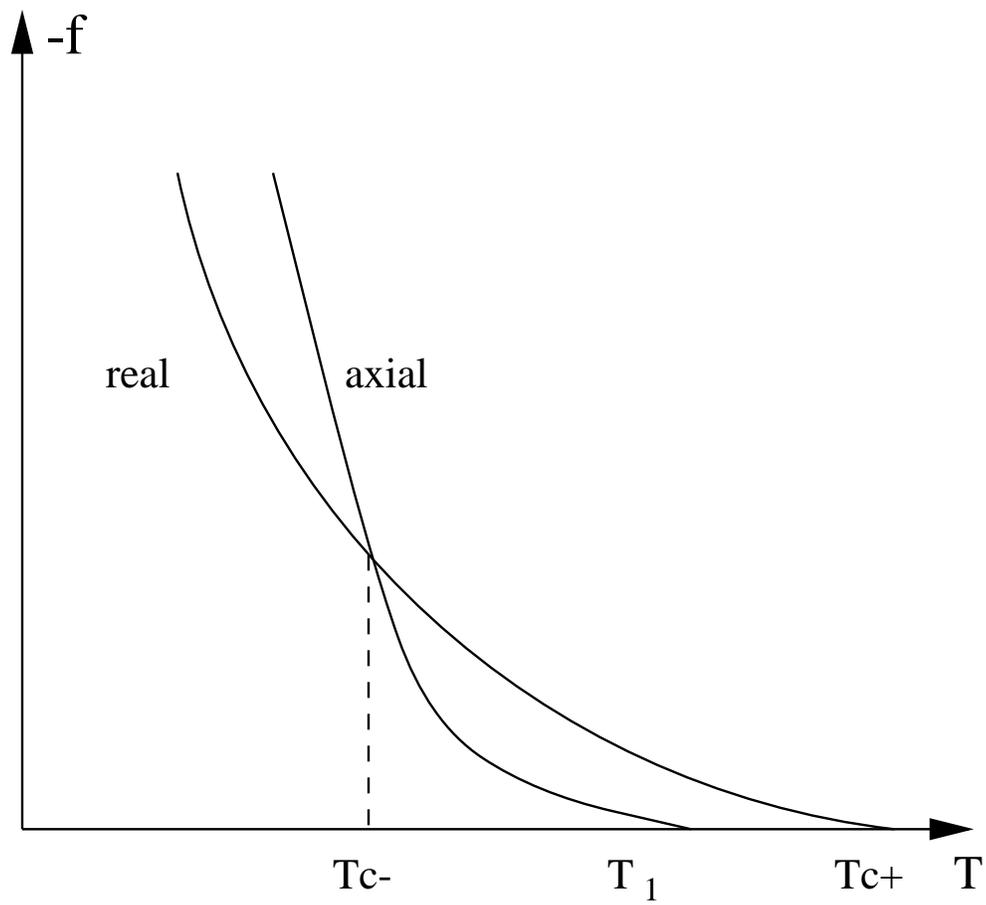
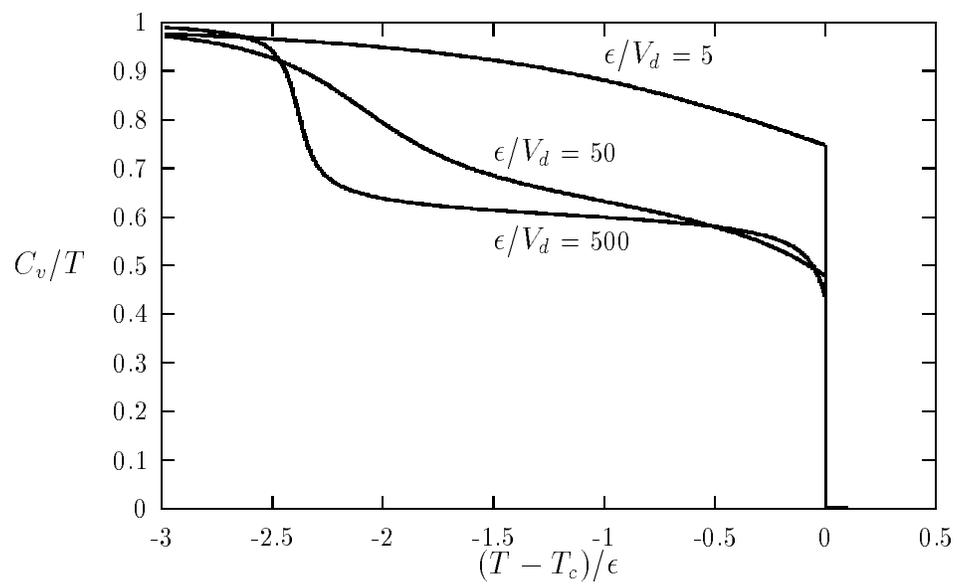
\end{document}